\documentclass[english,superscriptaddress,floatfix,twocolumn]{revtex4}
\usepackage[T1]{fontenc}
\usepackage[latin9]{inputenc}
\usepackage[pdftex]{color}
\usepackage{babel}
\usepackage{bm}
\usepackage{amsmath}
\usepackage{amssymb}
\usepackage[pdftex]{graphicx}
\usepackage{esint}

\makeatletter
\@ifundefined{textcolor}{}
{%
 \definecolor{BLACK}{gray}{0}
 \definecolor{WHITE}{gray}{1}
 \definecolor{RED}{rgb}{1,0,0}
 \definecolor{GREEN}{rgb}{0,1,0}
 \definecolor{BLUE}{rgb}{0,0,1}
 \definecolor{CYAN}{cmyk}{1,0,0,0}
 \definecolor{MAGENTA}{cmyk}{0,1,0,0}
 \definecolor{YELLOW}{cmyk}{0,0,1,0}
}

\usepackage{babel}
\usepackage{upgreek}
\usepackage{braket}

\makeatother

\begin{document}

\title{Lévy flights and hydrodynamic superdiffusion on the Dirac cone of
Graphene}

\author{Egor I. Kiselev}

\affiliation{Institut für Theorie der Kondensierten Materie, Karlsruher Institut
für Technologie, 76131 Karlsruhe, Germany}

\author{Jörg Schmalian}

\affiliation{Institut für Theorie der Kondensierten Materie, Karlsruher Institut
für Technologie, 76131 Karlsruhe, Germany}

\affiliation{Institut für Festkörperphysik, Karlsruher Institut für Technologie,
76131 Karlsruhe, Germany}
\begin{abstract}
We show that hydrodynamic collision processes of graphene at the neutrality
point can be described in terms of a Fokker-Planck equation with fractional
derivative, corresponding to a Lévy flight in momentum space. Thus,
electron-electron collisions give rise to frequent small-angle scattering
processes that are interrupted by rare large-angle events. The latter
give rise to superdiffusive dynamics of collective excitations. We
argue that such superdiffusive dynamics is of more general importance
to the out-of-equilibrium dynamics of quantum-critical systems.
\end{abstract}
\maketitle
The kinetics of large gravitational systems such as globular clusters
in galaxies or of a classical charged plasma are governed by continuous
collisions with small-angle scatterings. The origin for this behavior
is the long-range character of the Newton or Coulomb force, respectively.
Such small-angle collisions behave in velocity space like drag and
diffusion events, where a Fokker-Planck equation offers an efficient
description\cite{Chandrasekhar1943,Rosenbluth1957,Chernoff1990}.
Collisions can thus be seen as a Gaussian random walk in phase space.
The velocity of a plasma or gravitational dust particle undergoes
ordinary Brownian motion.

Quantum many-body systems that are near a quantum-critical point are
governed by soft modes that will also ind\textcolor{black}{uce effective
long-range interactions\cite{Sachdevbook}. This begs the question
whether such quantum-critical systems also allow for an effective
Fokker-Planck description of the non-equilibrium kinetics; in the
collision-dominated hydrodynamic regime and in the crossover regime
from hydrodynamic to ballistic dynamics. Candidate systems are itinerant
electrons near magnetic or nematic quantum phase transitions\cite{Hertz1976,Moriya1984,Millis1993,Laughlin2001,Millis2002,Abanov2003,Loehneysen2007,DellAnna2007,Metlitski2010,Schattner2016},
the superconductor-insulator phase transition\cite{Damle1997}, or
graphene near the Dirac point\cite{Sheehy2007}. Anomalous diffusion
was even shown to be present in two-dimensional Fermi liquids\cite{Gurzhi1995,Gurzhi1996,Maslov2011,Pal2012,Maslov2017,Ledwith2019}.}

\textcolor{black}{}
\begin{figure}
\textcolor{black}{\includegraphics[scale=0.4]{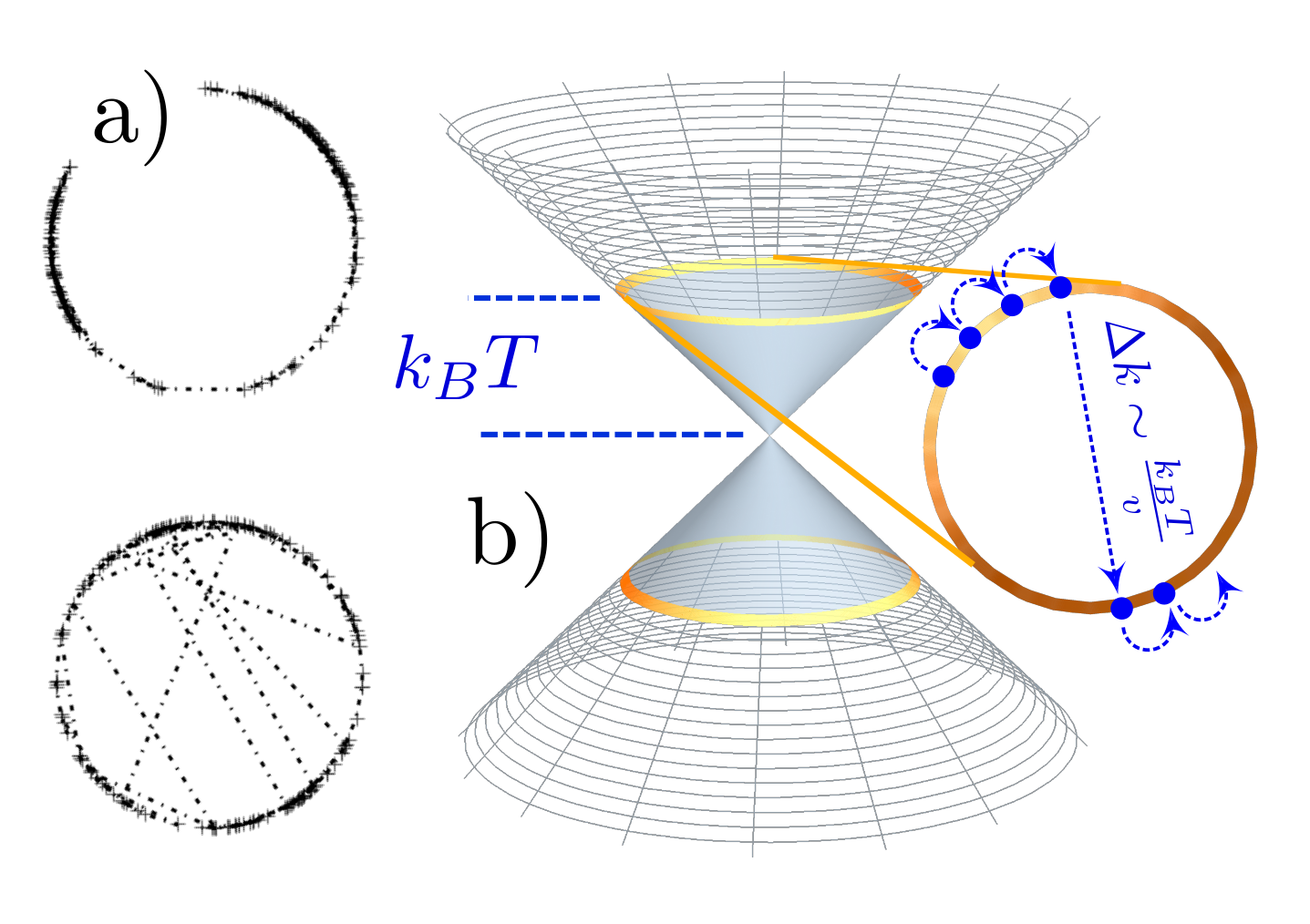}}

\textcolor{black}{\caption{a) A wrapped Gaussian flight (upper circle) and a wrapped Cauchy flight
(lower circle) with rare large momentum-transfer processes. b) Illustration
of the Lévy flight in momentum space for graphene at the Dirac point.
Electrons and holes that are thermally excited collide into each other.
Most of the time the momentum transfer due to the electron-electron
Coulomb interaction leads to small-angle scattering. However, those
processes are interrupted by rare processes with large momentum transfer.
The latter change the dynamics of the system qualitatively, leading
to an accelerated or superdiffusive dynamics.}
}

\textcolor{black}{\label{Fig: sketch}}
\end{figure}

\textcolor{black}{In this paper we analyze the quantum kinetics of
graphene near the Dirac point with electron-electron Coulomb interaction.
We show that the kinetic theory at charge neutrality\cite{Fritz2008,Kashuba2008,Mueller2009,Schuett2011,Kiselev2019}
can be expressed in terms of a Fokker-Planck equation, yet with fractional
derivative with respect to the momentum direction. The underlying
random processes are Lévy flights\cite{Levy1954,Mandelbrot1977,Feller1972},
non-Gaussian random walks whose step widths are distributed according
to a powerlaw. The slowly decaying tail of the step-width distribution
makes it impossible to define a diffusion constant or to use a conventional
Fokker-Planck equation. However, a diffusion equation of the form
\begin{equation}
\frac{\partial\rho}{\partial t}+D_{\mu}\left|\triangle\right|^{\frac{\mu}{2}}\rho=0,\label{eq:frac diffusion}
\end{equation}
with appropriately generalized fractional derivative\cite{Samko1993,Hermann2014}
can be used to describe such random walks. Lévy flights have been
discussed to model the migration pattern of animals as they search
for resources\cite{Bartumeus2005,Reynolds2009}, the high-frequency
index dynamics of the stock market\cite{Mantegna1997}, or to describe
durations between consecutive earthquakes\cite{Corral2006}. In our
system they correspond to random walks in momentum space with powerlaw
weight for large momentum-transfer processes. We demonstrate that
the collision operator due to electron-electron interactions in graphene
takes the form of a fractional derivative. Then the Boltzmann equation
becomes a fractional Fokker-Planck equation, similar to Eq.\ref{eq:frac diffusion}
with exponent $\mu=1$:
\begin{equation}
\left(\partial_{t}+\mathbf{v}_{\mathbf{k}\lambda}\cdot\nabla_{\mathbf{x}}-\tau_{L}^{-1}\left(\frac{\partial^{2}}{\partial\theta^{2}}\right)^{1/2}\right)f_{\mathbf{k}\lambda}=S_{\mathbf{k}\lambda},\label{eq:fractional FP}
\end{equation}
where $\theta$ determines the electron momentum direction: $\mathbf{k}=k\left(\cos\theta,\sin\theta\right)$.
The precise definition of the fractional derivative is given below.
This result implies that the out-of-equilibrium dynamics of graphene
in the hydrodynamic regime is governed by a wrapped Cauchy flight\cite{Levy1939,Mardia1999},
a specific Lévy flight on the Dirac cone. In Fig. \ref{Fig: sketch}a
we show a simulation of ordinary Brownian motion on a ring and of
the wrapped Cauchy flight. Details of this simulation are summarized
in\cite{supplementary material}. The occurrence of rare large-angle
jumps is clearly visible. The corresponding phase-space dynamics is
sketched in Fig. \ref{Fig: sketch}b. While the direction of $\mathbf{k}$
undergoes anomalous diffusion, its magnitude $k\equiv\left|\mathbf{k}\right|$
is of the order of $k_{B}T/v_{0}$ with the graphene group velocity
$v_{0}\approx10^{8}{\rm cm}/{\rm s}$. The characteristic time of
the process is $\tau_{L}$ with}

\textcolor{black}{
\begin{equation}
\hbar\tau_{L}^{-1}\approx11.66\alpha^{2}k_{B}T,\label{eq:Levy time}
\end{equation}
where the fine-structure constant of graphene is $\alpha=e^{2}/\left(\hbar\epsilon v_{0}\right)$.
$\tau_{L}$ agrees up to a numerical coefficient with the collision
time for the hydrodynamic transport behavior of graphene at the Dirac
point\cite{Fritz2008,Kashuba2008,Mueller2009}. Below we discuss how
$\tau_{L}$ is determined. Such a time scale was recently observed
experimentally in THz spectroscopy of graphene at charge neutrality\cite{Gallagher2019}. }

\textcolor{black}{Lévy flights in graphene have been discussed in
Ref.\cite{Gattenloehner2016}, where an egineered distribution of
adatoms was shown to result in a superdiffusive behavior of charge
carriers, and in Ref.\cite{Briskot2014} in the context of highly
photo-excited carriers that relax according to a cascade of processes
- a behavior with interesting implications for pump-probe experiments.
This can be seen as a superdiffusion in energy space far from equilibrium.
It affects the magnitude of the momentum. Here we focus on the low-energy
hydrodynamic regime and find a very different behavior for the directional
diffusion in momentum space. Nevertheless, these results strongly
suggest that superdiffusive phase-space dynamics is a more common
phenomenon in quantum-critical systems. }

\textcolor{black}{We start from the Boltzmann equation 
\begin{equation}
\left(\partial_{t}+\mathbf{v}_{\mathbf{k}\lambda}\cdot\nabla_{\mathbf{x}}+\mathbf{F}\left(\mathbf{x},t\right)\cdot\nabla_{\mathbf{k}}+{\cal C}\right)f_{\mathbf{k}\lambda}\left(\mathbf{x},t\right)=0\label{eq:Boltzmann}
\end{equation}
for the electron distribution function $f_{\mathbf{k}\lambda}\left(\mathbf{x},t\right)$
where $\mathbf{k}$ refers to the momentum and $\lambda=\pm1$ labels
the upper and lower cone of the Dirac spectrum $\varepsilon_{\mathbf{k}\lambda}=\lambda v_{0}\left|\mathbf{k}\right|$.
$\mathbf{v}_{\mathbf{k}\lambda}=\partial\varepsilon_{\mathbf{k}\lambda}/\partial\mathbf{k}$
is the velocity vector and $\mathbf{F}\left(\mathbf{x},t\right)$
some external force, e.g. due to an external electric field. ${\cal C}$
is the Boltzmann collision operator due to electron-electron interactions
and was derived to order $\alpha^{2}$ in Ref.\cite{Fritz2008} from
a Keldysh-Schwinger approach; see also in\cite{supplementary material}.
It takes the usual form of a two-body interaction:
\begin{eqnarray}
{\cal C}f_{1} & = & -\sum_{2,3,4}W_{12,34}\left[f_{1}f_{2}\left(1-f_{3}\right)\left(1-f_{4}\right)\right|\nonumber \\
 & - & \left|\left(1-f_{1}\right)\left(1-f_{2}\right)f_{3}f_{4}\right].
\end{eqnarray}
The transition probability $W_{12,34}$ is due to the electron-electron
Coulomb interaction $e^{2}/\epsilon$ of Dirac fermions that are confined
to a two-dimensional system. $\epsilon$ is the dielectric constant
determined by the substrate. For free standing graphene $\epsilon=1$
and the fine-structure constant $\alpha\approx2.2$ is of order unity.
A renormalization group analysis shows that $\alpha$ flows towards
weak coupling, justifying our perturbative approach\cite{Sheehy2007}.}

\textcolor{black}{As usual, the kinetic distribution function $f_{\lambda,\bm{k}}$
is expanded around the local equilibrium distribution $f_{k\lambda}^{0}=\left(e^{\beta\left(\epsilon_{\lambda,\bm{k}}-\mu\right)}+1\right)^{-1}$
and parametrized as ($f_{k}^{\left(0\right)}=f_{k+}^{\left(0\right)}$):
\begin{equation}
f_{\mathbf{k}\lambda}\left(\mathbf{x},t\right)=f_{k\lambda}^{\left(0\right)}+f_{k}^{\left(0\right)}\left(1-f_{k}^{\left(0\right)}\right)\psi_{\mathbf{k}\lambda}\left(\mathbf{x},t\right).
\end{equation}
We linearize the Boltzmann equation with respect to $\psi_{\mathbf{k}\lambda}\left(\mathbf{x},t\right)$.
With the Liouville operator
\begin{equation}
{\cal L}=\left(\partial_{t}+\mathbf{v}_{\mathbf{k}\lambda}\cdot\nabla_{\mathbf{x}}\right)f_{k}^{\left(0\right)}\left(1-f_{k}^{\left(0\right)}\right)
\end{equation}
we obtain a compact formulation of the Boltzmann equation: $\left({\cal L}+{\cal C}\right)\psi=S$.
$S_{\mathbf{k}\lambda}\left(\mathbf{x},t\right)$ contains external
perturbations, such as those due to a space and time dependent electric
field or flow-velocity gradient. The operators ${\cal L}$ and ${\cal C}$
act on the momentum and band indices $\mathbf{k}$ and $\lambda$,
respectively. Taking into account the kinematic constraints of the
linear Dirac spectrum, the collision operator becomes:}

\textcolor{black}{
\begin{eqnarray}
\left({\cal C}\psi\right)_{\mathbf{k}\lambda} & = & \frac{2\pi}{\hbar}\int_{k^{\prime}q}\delta\left(k+k'-\left|\mathbf{k}+\mathbf{q}\right|-\left|\mathbf{k}'-\mathbf{q}\right|\right)\\
 & \times & \left(1-f_{k}^{\left(0\right)}\right)\left(1-f_{k'}^{\left(0\right)}\right)f_{\left\vert \mathbf{k}+\mathbf{q}\right\vert }^{\left(0\right)}f_{\left\vert \mathbf{k}'-\mathbf{q}\right\vert }^{\left(0\right)}\nonumber \\
 & \times & \left\{ \gamma_{\mathbf{k},\mathbf{k}^{\prime},\mathbf{q}}^{\left(1\right)}\left(\psi_{\mathbf{k}+\mathbf{q}\lambda}+\psi_{\mathbf{k}'-\mathbf{q}\lambda}-\psi_{\mathbf{k}'\lambda}-\psi_{\mathbf{k}\lambda}\right)\right.\nonumber \\
 & + & \left.\gamma_{\mathbf{k},\mathbf{k}^{\prime},\mathbf{q}}^{\left(2\right)}\left(\psi_{\mathbf{k}+\mathbf{q}\lambda}-\psi_{-\mathbf{k}'+\mathbf{q}\bar{\lambda}}+\psi_{-\mathbf{k}'\bar{\lambda}}-\psi_{\mathbf{k}\lambda}\right)\right\} ,\nonumber 
\end{eqnarray}
where the matrix elements $\gamma_{\mathbf{k},\mathbf{k}',\mathbf{q}}^{\left(1,2\right)}$
are given in Ref.\cite{supplementary material} and $\int_{k}\cdots=\int\frac{d^{2}k}{\left(2\pi\right)^{2}}\cdots$.
One easily finds the zero modes that correspond to the conservation
laws\cite{Fritz2008}. Eq.\ref{eq:Boltzmann} was obtained by projecting
the distribution function onto the helical eigenstates of the problem.
The same projection was performed in the derivation of the collision
operator\cite{Fritz2008,supplementary material}.}

\textcolor{black}{The usual analysis of the Boltzmann equation proceeds
as follows: One performs a Fourier transformation from $\left(\mathbf{x},t\right)$
to $\left(\mathbf{q},\omega\right)$ and introduces a complete set
of states $\chi_{\mathbf{k}\lambda}^{\left(s\right)}$ to evaluate
the matrix elements $\left\langle s\left|{\cal L}+{\cal C}\right|s'\right\rangle $
with scalar product $\left\langle s|s'\right\rangle =\sum_{\mathbf{k}\lambda}\chi_{\mathbf{k}\lambda}^{\left(s\right)*}\chi_{\mathbf{k}\lambda}^{\left(s'\right)}$.
The Liouville operator becomes ${\cal L}=\left(-i\omega+i\mathbf{v}_{\mathbf{k}\lambda}\cdot\mathbf{q}\right)f_{k}^{\left(0\right)}\left(1-f_{k}^{\left(0\right)}\right)$.
The distribution function then follows as $\psi=\left({\cal L}+{\cal C}\right)^{-1}S$.
For finite $\omega$ or $\mathbf{q}$ the operator ${\cal L}+{\cal {\cal C}}$
is nonsingular. This program is somewhat simplified for graphene at
charge neutrality. As shown in Refs.\cite{Fritz2008,Kashuba2008,Mueller2009,Kiselev2019},
scattering processes where all momenta are collinear are enhanced
by a factor $\log\left(1/\alpha\right)$. This can be used to identify
the dominant modes, derived in the supplementary material:
\begin{equation}
\chi_{\mathbf{k}\lambda}^{\left(m,s\right)}=\lambda^{m}e^{im\theta}\left\{ 1,\lambda,\lambda v_{0}k/\left(k_{B}T\right)\right\} ,\label{eq:Collinear_modes}
\end{equation}
where $m\in\mathbb{Z}$ is the angular momentum quantum number while
$s=1\cdots$3 labels the collinear modes for given $m$. We solve
the kinetic equation by projecting it onto the dominant collinear
modes $\chi_{\mathbf{k}\lambda}^{\left(m,s\right)}$, but checked
that our key conclusions are unchanged if we chose a larger set of
basis functions. Also, if we restrict our considerations to the transport
of charge due to external electric fields, it suffices to consider
the modes $\chi_{\mathbf{k}\lambda}^{\left(m,1\right)}=\lambda^{m}e^{im\theta}$
of Eq. (\ref{eq:Collinear_modes}). For simplicity we confine ourselves
to electric-field source terms and only discuss this mode. The generalization
to other modes is straightforward.}

\textcolor{black}{The low-energy Dirac Hamiltonian is rotationally
invariant such that the collision operator becomes diagonal in the
angular momentum representation
\begin{equation}
\left\langle m\left|{\cal C}\right|m'\right\rangle =\frac{\ln2}{\pi}\delta_{m,m'}\tau_{m}^{-1}.
\end{equation}
The diagonal elements are, besides a convenient prefactor, the scattering
rates of the corresponding angular momentum channel. $\tau_{0}^{-1}=0$
due to charge conservation, while the collision rate 
\begin{equation}
\hbar\tau_{1}^{-1}=3.646\alpha^{2}k_{B}T\label{eq:cll rate tau1}
\end{equation}
for $m=1$ was determined in Ref.\cite{Fritz2008} to yield the optical
conductivity $\sigma\left(\omega\right)=\frac{e^{2}}{h}4\ln2k_{B}T\left(-i\hbar\omega+\hbar\tau_{1}^{-1}\right)^{-1}.$
$\tau_{1}^{-1}$ was recently observed in Ref.\cite{Gallagher2019}
using a waveguide setup; a demonstration of quantum-critical hydrodynamic
transport. The dramatic violation of the Wiedemann-Franz law at charge
neutrality is another important indication for electronic hydrodynamics
at charge neutrality\cite{Crossno2016}. }

\begin{figure}
\textcolor{black}{\includegraphics[scale=0.32]{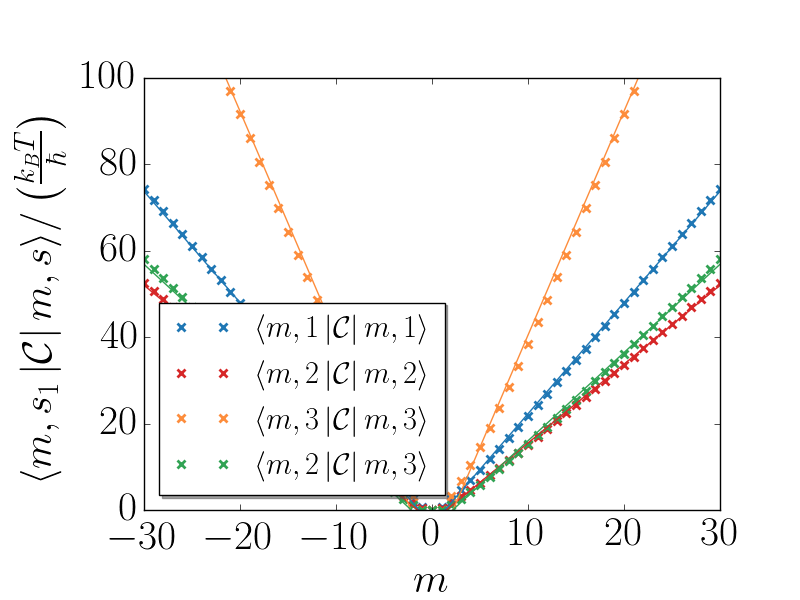}}

\textcolor{black}{\includegraphics[scale=0.32]{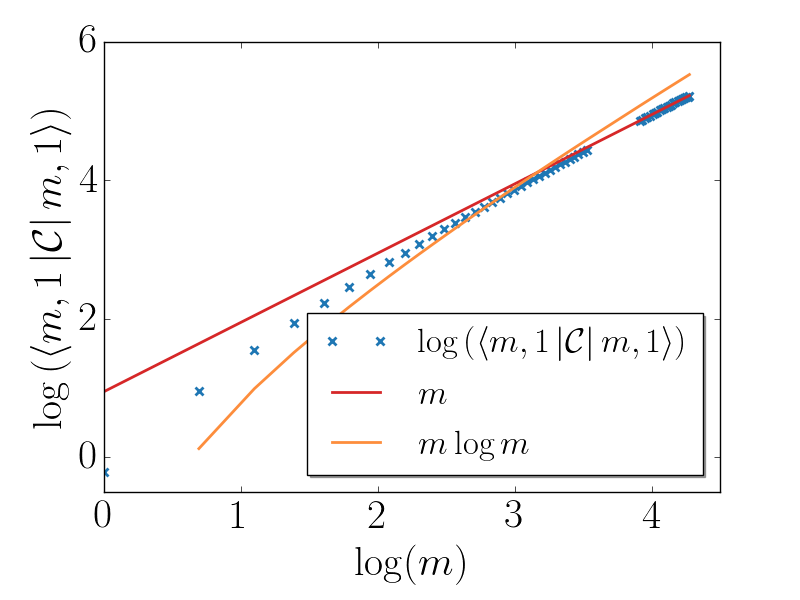}}

\textcolor{black}{\caption{Upper panel: Angular momentum dependence of the matrix elements of
the collision operator $\left\langle m,s\left|{\cal C}\right|m,s'\right\rangle $
where $s=1\cdots3$ refers to the collinear eigenmodes of Eq.\ref{eq:Collinear_modes}.
In the text we discuss, for simplicity, only $\left\langle m\left|{\cal C}\right|m\right\rangle \equiv\left\langle m,1\left|{\cal C}\right|m,1\right\rangle $.
Lower panel: log-log plot of the matrix element to demonstrate that
we can distinguish the $\left|m\right|$-dependence from, e.g. $\left|m\right|\log\left|m\right|$.}
}

\textcolor{black}{\label{Fig: m dependence}}
\end{figure}

\textcolor{black}{We evaluated the matrix elements $\left\langle m\left|{\cal C}\right|m\right\rangle $
and obtain
\begin{equation}
\tau_{m}^{-1}=\tau_{1}^{-1}\left(\kappa\left|m\right|-\kappa'\right),
\end{equation}
where the two numerical constants are given as $\kappa\approx3.199$
and $\kappa'\approx4.296$, see also Fig. \ref{Fig: m dependence}.
This behavior is asymptotically exact at large $m$ but valid with
good accuracy already for $m>2$. The most important aspect of this
result is that the dependence of the scattering rate on the angular
momentum $m$ is non-analytic. To simplify the analysis we assume
in the following that $\tau_{m}^{-1}=\tau_{L}^{-1}\left|m\right|$,
where $\tau_{L}^{-1}=\kappa\tau_{1}^{-1}$ is the characteristic time
of the of the Lévy flight process, given in Eq. (\ref{eq:Levy time}).}

\textcolor{black}{}

\textcolor{black}{The implication of the $\left|m\right|$-dependence
of $\tau_{m}^{-1}$ becomes evident if we consider the scattering
between two distinct momentum directions. Fourier transformation of
$\tau_{m}^{-1}$ yields:
\begin{eqnarray}
\left\langle \theta\left|{\cal C}\right|\theta'\right\rangle  & = & -\frac{\ln2\tau_{L}^{-1}}{\left(2\pi\right)^{2}\sin^{2}\left(\frac{\theta-\theta'}{2}\right)}.\label{eq:coll theta}
\end{eqnarray}
Thus, we obtain a slowly-decaying powerlaw $\sim\left(\theta-\theta'\right)^{-2}$
for scattering processes away from forward scattering. Using this
result for $\left\langle \theta\left|{\cal C}\right|\theta'\right\rangle $
we can rewrite the Boltzmann equation in the form Eq. (\ref{eq:fractional FP})
with characteristic time $\tau_{L}$ of Eq. (\ref{eq:Levy time})
for the Lévy flight. To arrive at Eq. (\ref{eq:fractional FP}) we
used that the convolution of the distribution function with $\left\langle \theta\left|{\cal C}\right|\theta'\right\rangle $
can be expressed as a fractional derivative 
\begin{equation}
\left(\frac{\partial^{2}g\left(\theta\right)}{\partial\theta^{2}}\right)^{1/2}=\frac{\partial}{\partial\theta}\int_{0}^{2\pi}\frac{g\left(\theta'\right)}{\tan\left(\frac{\theta-\theta'}{2}\right)}d\theta',
\end{equation}
 a special case of the Riesz-Feller derivative $\triangle^{\mu/2}$\cite{Samko1993,Hermann2014}. }

\begin{figure}
\begin{centering}
\textcolor{black}{\includegraphics[scale=0.32]{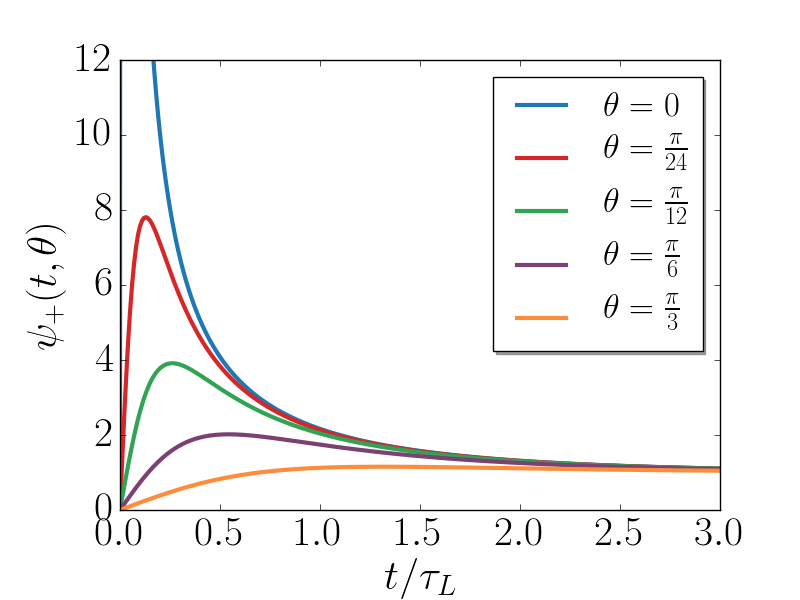}}
\par\end{centering}

\begin{centering}
\textcolor{black}{\includegraphics[scale=0.32]{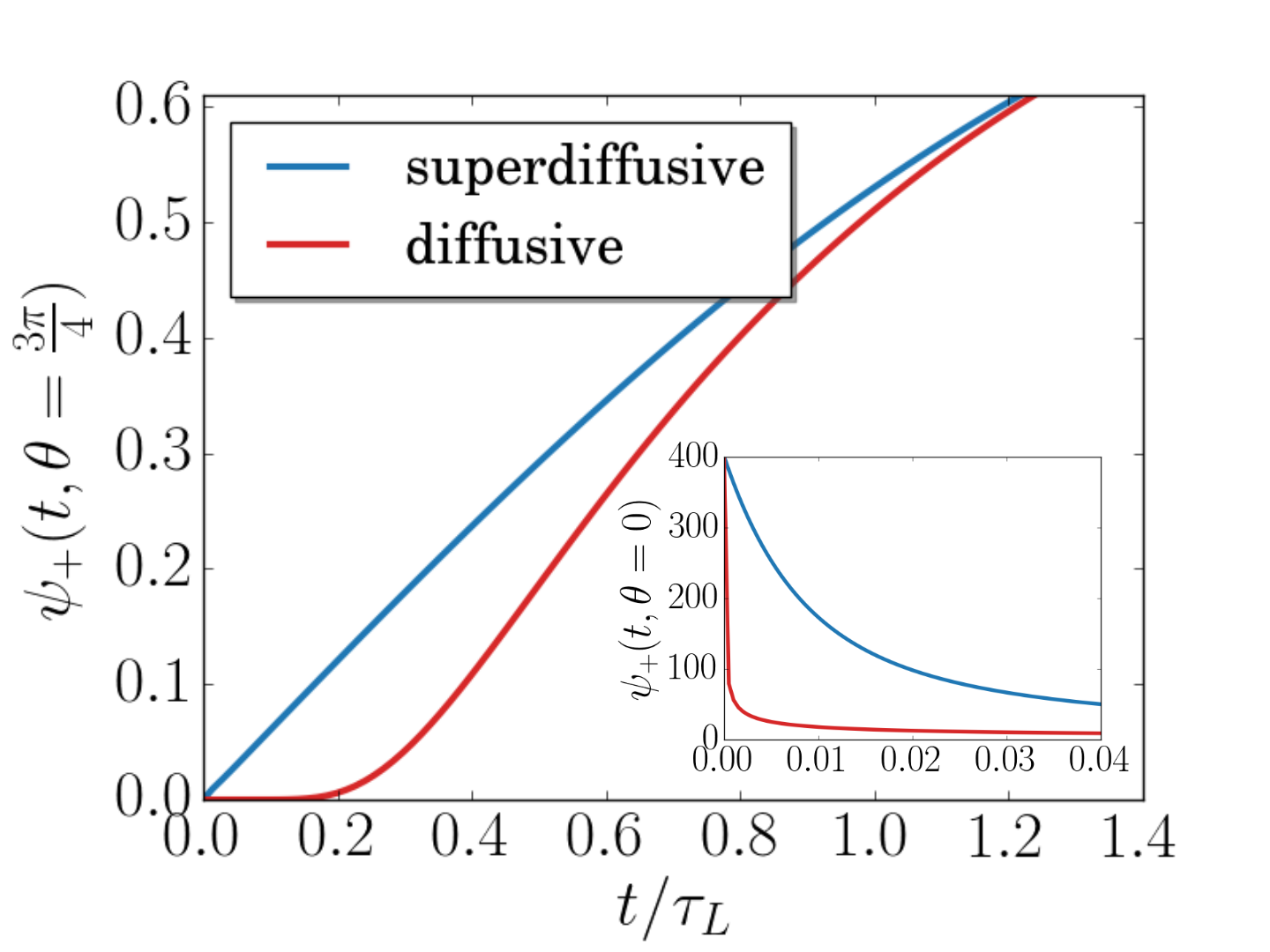}}
\par\end{centering}

\textcolor{black}{\caption{Upper panel: Post-injection distribution function that follows from
the fractional Fokker-Planck equation, Eq.\ref{eq:fractional FP},
with external perturbation of Eq.\ref{eq:injection}. Notice the superdiffusive
dynamics at short times. Lower panel: Comparison of superdiffusive
and diffusive dynamics at short times. At angles away from the peak
at $\theta=0$ superdiffusion leads to a faster growth of the distribution
function. Inlet: the initial peak at $\theta=0$ decays as $1/t$
for superdiffusion and $1/\sqrt{t}$ for ordinary diffusion. This
behavior dominates the heating of the system (see main text).}
}

\textcolor{black}{\label{Fig: distribution fct}}
\end{figure}

\textcolor{black}{There are some profound implications that this fractional
Fokker-Planck formulation immediately reveals. For example, we consider
a scenario where we inject a highly directed excitation\cite{Ledwith2019}.
To this end we consider a source term in the Boltzmann equation that
causes this excitation:
\begin{equation}
S_{\mathbf{k}\lambda}\left(t\right)=\delta\left(t\right)f_{k}^{\left(0\right)}\left(1-f_{k}^{\left(0\right)}\right)\sum_{m}\delta h_{\lambda m}e^{im\theta}.\label{eq:injection}
\end{equation}
We assumed that we will only inject excitations in a window $\pm k_{B}T$
near the Dirac point, hence the factor $f_{k}^{\left(0\right)}\left(1-f_{k}^{\left(0\right)}\right)$.
In addition we decomposed the source term into its angular momentum
modes. The linearized Boltzmann equation is applicable if $\left|\delta h_{\lambda m}\right|\ll1$.
To describe an excitation that is peaked along an axis given by a
certain momentum direction, we use $\delta h_{\lambda,m}=\delta h\lambda^{m}$,
which has a $\lambda$-dependence of the $s=1$ mode of Eq. (\ref{eq:Collinear_modes}).
The solution of the fractional Fokker-Planck equation for a homogeneous
case $\mathbf{q}=0$ is then given as 
\begin{equation}
\psi_{\lambda}\left(\theta,t\right)=\delta h\Theta\left(t\right)\frac{\sinh\left(t/\tau_{L}\right)}{\cosh\left(t/\tau_{L}\right)-\lambda\cos\left(\theta\right)}.\label{eq:Distribution function}
\end{equation}
This function is known as wrapped Cauchy distribution with circular
variance $1-e^{-t/\tau_{L}}$\cite{Levy1939,Mardia1999}. $\Theta\left(t\right)$
is the step function. $\psi_{+}\left(\theta,t\right)$ is shown in
the upper panel of Fig. \ref{Fig: distribution fct}.}

\textcolor{black}{}

\textcolor{black}{For $t=0$, $\psi_{\lambda}\left(\theta,t\right)$
corresponds to two delta functions due to particle and hole flows
in opposite directions. Let us concentrate on the particle channel
$\lambda=+1$. For short times $t\ll\tau_{L}$, the peak in the initial
current direction decays as
\begin{equation}
\psi_{+}\left(t,\theta=0\right)\approx\delta h\frac{\tau_{L}}{\pi t},
\end{equation}
while the distribution function grows linearly for all non-zero angles:
\begin{equation}
\psi_{+}\left(t,\theta\neq0\right)\approx\frac{\delta h}{4\pi\sin^{2}\left(\theta/2\right)}\frac{t}{\tau_{L}}.
\end{equation}
The same behavior occurs for $\lambda=-1$ if we shift $\theta\rightarrow\theta+\pi$.
This behavior in contrast to the one that follows from usual Fokker-Planck
diffusion. The latter we obtain for example from collision rates $\tau_{m}^{-1}\sim m^{2}.$
Then the usual spreading of a Gaussian wave package occurs with $\psi_{+}\left(t,\theta=0\right)\propto t^{-1/2}$
and $\psi_{+}\left(t,\theta\neq0\right)\propto t^{2}$ (lower panel
of Fig. \ref{Fig: distribution fct}). While the forward direction
of a Levy flight decays more slowly than in usual diffusion, the growth
at larger angles is much faster, hence the name superdiffusion. }

\textcolor{black}{A tangible implication of this superdiffusive charge
motion is the heating of the system after the injection. To this end
we determine the time dependence of the entropy density 
\begin{equation}
\frac{\partial s\left(t\right)}{\partial t}=4k_{B}\sum_{\lambda}\int_{k}\log\left(\frac{1-f_{\mathbf{k}\lambda}}{f_{\mathbf{k}\lambda}}\right)\frac{\partial f_{\mathbf{k\lambda}}}{\partial t}.
\end{equation}
The heat density caused by the injection is given by $\delta q\left(t\right)=T\left(s_{{\rm eq}}-s\left(t\right)\right).$
Inserting the distribution function of Eq. (\ref{eq:Distribution function})
we obtain
\begin{equation}
\frac{\partial s\left(t\right)}{\partial t}=\frac{4\log2}{9\zeta\left(3\right)}\frac{s_{{\rm eq}}}{\tau_{L}}\frac{\left(\delta h\right)^{2}}{\sinh^{2}\left(t/\tau_{L}\right)},
\end{equation}
where $s_{{\rm eq}}$ is the equilibrium entropy density. In order
to stay within the regime of linear response, we are confined to $t>\delta h\tau_{L}$.
For $t\rightarrow\infty$ one finds $s\rightarrow s_{{\rm eq}},$
and we obtain $s\left(t\right)=s_{{\rm eq}}\left(1-\left(\delta h\right)^{2}\frac{4\log(2)}{9\zeta\left(3\right)}\left(\coth\left(\frac{t}{\tau_{L}}\right)-1\right)\right)$.
Thus, initial heating occurs according to 
\begin{equation}
\delta q\left(t\right)\propto Ts_{{\rm eq}}\delta h^{2}\frac{\tau_{L}}{t}.\label{eq:short-time-heating}
\end{equation}
This result is a direct consequence of the superdiffusive behavior,
in particular of the slow decay along the forward direction. In case
of ordinary diffusion follows $\delta q\left(t\right)\propto t^{-1/2}$
which is much faster (see Fig. \ref{Fig: distribution fct}). The
$m$-dependence of $\tau_{m}^{-1}$ that is responsible for the Lévy
flight behavior can also be seen in non-local transport coefficients
since the conductivity at finite momentum $\mathbf{q}$ couples the
different harmonics of the distribution function. As an example we
show in the supplementary material the transverse optical conductivity
at finite $\mathbf{q}$. Nevertheless, experiments with directed electron
beams \cite{Wang2019}, which in the past have been used to investigate
electron-electron scattering effects \cite{Predel2000}, seem to offer
a more direct way of testing the short time behavior of Eq. (\ref{eq:short-time-heating}).}

\textcolor{black}{The occurrence of Lévy flights to describe scattering
processes in momentum space is a more general phenomenon and not restricted
to graphene at the neutrality point. In two-dimensional Fermi liquids
with characteristic rate $\hbar\tau_{FL}^{-1}\sim k_{B}T^{2}/T_{F}$,
it holds for $\left|m\right|<M\sim\sqrt{T_{F}/T}$ that $\tau_{m}$$^{-1}\sim\tau_{FL}^{-1}\frac{m^{p}}{M^{p}}\log\left|m\right|$
with $p=2\left(1+\left(-1\right)^{m}\right)$, while $\tau_{m}^{-1}\sim\tau_{FL}^{-1}$
for $\left|m\right|>M$\cite{Gurzhi1996,Ledwith2019}. $T_{F}$ is
the Fermi temperature. This yields superdiffusive behavior in a wide
time window. Another system that also shows $\tau_{m}^{-1}\propto\left|m\right|$
for arbitrarily large $m$ consists of electrons in a random magnetic
field, important for the description of composite fermions in the
fractional quantum Hall regime\cite{Mirlin1997}. Our analysis implies
that this system should also undergo a wrapped Cauchy flight in momentum
space. Large classes of quantum-critical systems, discussed e.g. in
Refs.\cite{Hertz1976,Moriya1984,Millis1993,Laughlin2001,Millis2002,Abanov2003,Loehneysen2007,DellAnna2007,Metlitski2010,Schattner2016,Damle1997}
are governed by long-ranged soft-mode interactions. An analysis of
collision processes along the lines discussed here may reveal a non-analytic
dependence of the scattering rates on angular momentum quantum number
according to $\tau_{m}^{-1}\propto\left|m\right|^{\mu/2}$. This would
give rise to a more general class of wrapped Lévy flights, a consequence
of the power-law behavior $\left\langle \theta\left|{\cal C}\right|\theta'\right\rangle \propto\left|\theta-\theta'\right|^{-1-\frac{\mu}{2}}$
near forward scattering. This could occur on the Fermi surface for
itinerant quantum critical systems or near a soft momentum in critical
bosonic systems. If a fractional Fokker-Planck formulation, along
the lines of our Eq. (\ref{eq:fractional FP}), can be derived, it
will be significantly easier to draw conclusions about the out-of-equilibrium
dynamics of the system such as a focussed injection of collective
excitations. Finally we mention that the formulation of the Boltzmann
equation presented here can also be used to study the non-local electric
and thermal conductivities and viscosities, allowing insight into
the diffusive and sound excitations in the hydrodynamic regime\cite{Kiselev2019b}.}

\textcolor{black}{Acknowledgements: We are grateful to Andrey V. Chubukov,
Leonid S. Levitov, Dimitrii L. Maslov, and Alexander D. Mirlin for
stim}ulating discussions and to the European Commission's Horizon
2020 RISE program \emph{Hydrotronics} for support.

\pagebreak

\onecolumngrid

\part*{Supplementary material}

\section{Summary of the Simulations shown in Fig.1}

\begin{figure}
\centering{}\includegraphics[scale=0.35]{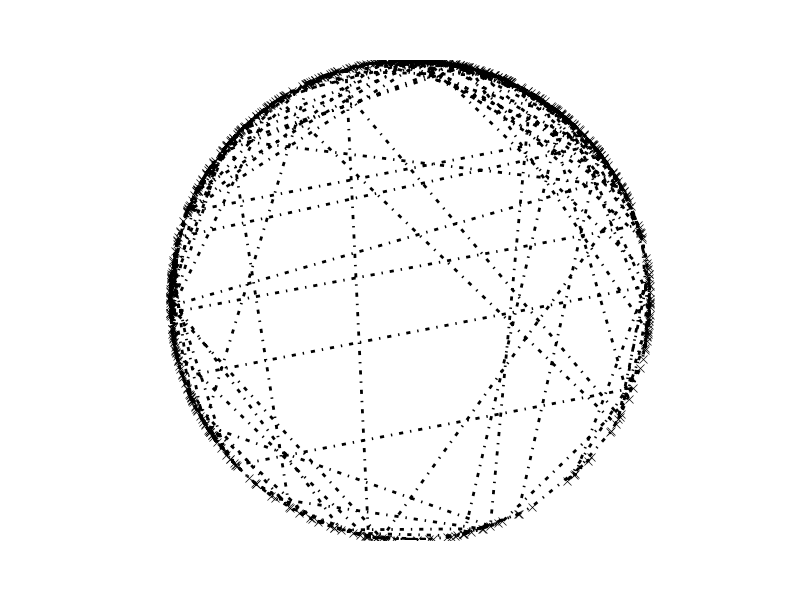}
\caption{500 steps of a superdiffusive wrapped Cauchy random walk of an electron
on the Dirac cone.\label{fig:Superdiffusive-wrapped-Cauchy}}
\end{figure}

\begin{figure}
\centering{}\includegraphics[scale=0.35]{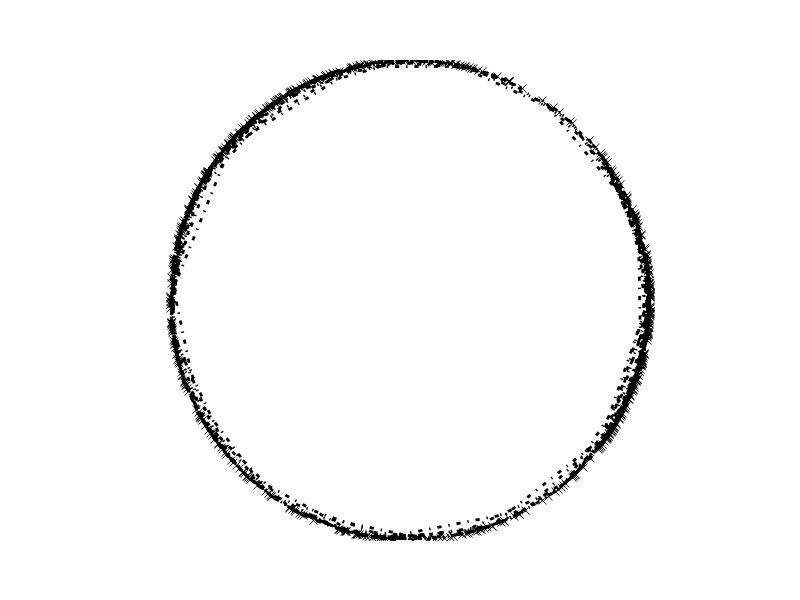}\caption{500 steps of a wrapped Gaussian random walk. The wrapped normal distribution
was approximated by the von Mises distribution.\label{fig:500-steps-of_gaussian} }
\end{figure}

Superdiffusion on the Dirac cone can be seen as a random walk of particles,
where the step sizes are distributed according to a wrapped Cauchy
distribution. This distribution solves the fractional Fokker-Planck
equation (2) of the main text (see \cite{Chechkin2004,Metzler2012}).
The anglular distance on the Dirac cone travelled by an electron during
a time interval $\Delta t$ is therefore distributed according to
\begin{equation}
\tilde{\psi}\left(\theta,\Delta t\right)=\frac{\sinh\left(\Delta t/\tau_{L}\right)}{\cosh\left(\Delta t/\tau_{L}\right)-\cos\left(\theta\right)}.\label{eq:Wrapped_Cauchy}
\end{equation}
To generate Figs. 1 a) and b) of the main text we created a sequence
of random angles $\Delta\theta_{i}$ using the distribution (\ref{eq:Wrapped_Cauchy}).
The position of the particle after $N$ steps, i.e. after a time intervall
$N\Delta t$, then is 
\begin{equation}
\theta_{N}=\sum_{i=0}^{N}\Delta\theta_{i}.
\end{equation}
Fig. \ref{fig:Superdiffusive-wrapped-Cauchy} depicts a wrapped Cauchy
random walk with $N=500$ steps. 
In the case of ordinary diffusion, the step size distribution of Eq.
(\ref{eq:Wrapped_Cauchy}) must be replaced by a wrapped normal distribution,
which is written in terms of the Jacobi theta function, but can be
closely approximated by the van Mises distribution (see e.g. Ref.
\cite{Kurz2014}): 
\[
\tilde{\psi}_{\textrm{normal}}\left(\theta,\Delta t\right)=\frac{e^{\cos\left(\theta\right)/\Delta t}}{2\pi I_{0}\left(1/\Delta t\right)}.
\]
Using the described procedure we obtain the wrapped Gaussian random
flight shown in Fig. \ref{fig:500-steps-of_gaussian}.

\section{Collision operator due to electron-electron Coulomb interaction}

We briefly summarize the main steps in deriving the collision operator
of the Boltzmann equation used in this paper. The collision operator
is determined from the larger and smaller self energies on the Keldysh
contour. For further details, see Ref. \cite{Fritz2008-1}.

The non-interacting part of the Hamiltonian is 
\begin{equation}
H_{0}=v\hbar\int_{k}\sum_{\alpha\beta}\psi_{\alpha}^{\dagger}\left(\mathbf{k}\right)\left(\mathbf{k\cdot\sigma}\right)_{\alpha\beta}\psi_{\beta}\left(\mathbf{k}\right)
\end{equation}
which is diagonalized by the unitary transformation 
\begin{equation}
U_{\mathbf{k}}=\frac{1}{\sqrt{2}}\left(\begin{array}{cc}
\frac{k_{x}+ik_{y}}{k} & 1\\
-\frac{k_{x}+ik_{y}}{k} & 1
\end{array}\right)
\end{equation}
 with
\begin{equation}
U_{\mathbf{k}}v\hbar\mathbf{k\cdot\sigma}U_{\mathbf{k}}^{-1}=\left(\begin{array}{cc}
v\hbar k & 0\\
0 & -v\hbar k
\end{array}\right).
\end{equation}
The eigenvalues of the Hamiltonian are $\pm v\hbar k$. Thus we obtain
quasiparticle states for the two bands: $\gamma_{\mathbf{k}}=U_{\mathbf{k}}\psi_{\mathbf{k}}$
with

\begin{equation}
H_{0}=v\hbar\int_{\mathbf{k}}\sum_{\lambda=\pm}\lambda k\gamma_{\mathbf{k},\lambda}^{\dagger}\gamma_{\mathbf{k},\lambda}.
\end{equation}
The electron-electron Coulomb interaction is 
\begin{equation}
H_{{\rm int}}=\frac{1}{2}\int_{k,k^{\prime},q}\sum_{\alpha\beta}V\left(\mathbf{q}\right)\psi_{\alpha}^{\dagger}\left(\mathbf{k+q},t\right)\psi_{\beta}^{\dagger}\left(\mathbf{k}^{\prime}-\mathbf{q},t\right)\psi_{\beta}\left(\mathbf{k}^{\prime},t\right)\psi_{\alpha}\left(\mathbf{k,}t\right)
\end{equation}
with $V\left(q\right)=\frac{e^{2}}{2\pi\varepsilon\left\vert \mathbf{q}\right\vert }$.
Transforming the interaction into the band, or helical representation,
which takes into account the locking between momentum and pseudo-spin
that originates from the two sub-lattice structure of graphene. It
follows 
\begin{equation}
H_{{\rm int}}=\frac{1}{2}\int_{k,k^{\prime},q}\sum_{\alpha\beta}T_{\lambda\mu\mu^{\prime}\lambda^{\prime}}\left(\mathbf{k,k}^{\prime},\mathbf{q}\right)\gamma_{\lambda^{\prime}}^{\dagger}\left(\mathbf{k+q},t\right)\gamma_{\mu}^{\dagger}\left(\mathbf{k}^{\prime}-\mathbf{q},t\right)\gamma_{\mu^{\prime}}\left(\mathbf{k}^{\prime},t\right)\gamma_{\lambda}\left(\mathbf{k,}t\right)
\end{equation}
where
\begin{equation}
T_{\lambda\mu\mu^{\prime}\lambda^{\prime}}\left(\mathbf{k,k}^{\prime},\mathbf{q}\right)=V\left(q\right)\left(U_{\mathbf{k+q}}U_{\mathbf{k}}^{-1}\right)_{\lambda^{\prime}\lambda}\left(U_{\mathbf{k}^{\prime}-\mathbf{q}}U_{\mathbf{k}^{\prime}}^{-1}\right)_{\mu\mu^{\prime}}.
\end{equation}

Within second order perturbation theory it holds for the self energies
for occupied and unoccupied states, respectively. 
\begin{eqnarray}
\Sigma{}_{\lambda}^{\gtrless}\left(\mathbf{k,}\omega\right) & = & N\sum_{\mu\mu^{\prime}\lambda^{\prime}}\int\frac{d^{2}qd^{2}k^{\prime}d\omega_{1}d\omega_{2}}{\left(2\pi\right)^{6}}\left\vert T_{\lambda\mu\mu^{\prime}\lambda^{\prime}}\left(\mathbf{k,k}^{\prime},\mathbf{q}\right)\right\vert ^{2}\nonumber \\
 & \times & G_{\lambda^{\prime}}^{\gtrless}\left(\mathbf{k+q,}\omega_{1}\right)G_{\mu}^{\gtrless}\left(\mathbf{k}^{\prime}-\mathbf{q,}\omega_{2}\right)G_{\mu^{\prime}}^{\lessgtr}\left(\mathbf{k}^{\prime}\mathbf{,}\omega_{1}+\omega_{2}-\omega\right)\nonumber \\
 & - & \sum_{\mu\mu^{\prime}\lambda^{\prime}}\int\frac{d^{2}qd^{2}k^{\prime}}{\left(2\pi\right)^{4}}\int\frac{d\omega_{1}d\omega_{2}}{\left(2\pi\right)^{2}}T_{\lambda\lambda^{\prime}\mu^{\prime}\mu}\left(\mathbf{k,k}^{\prime},\mathbf{k}^{\prime}-\mathbf{q-k}\right)T_{\lambda\mu\mu^{\prime}\lambda^{\prime}}\left(\mathbf{k,k}^{\prime},\mathbf{q}\right)^{\ast}\nonumber \\
 & \times & G_{\lambda^{\prime}}^{\gtrless}\left(\mathbf{k+q,}\omega_{1}\right)G_{\mu}^{\gtrless}\left(\mathbf{k}^{\prime}-\mathbf{q,}\omega_{2}\right)G_{\mu^{\prime}}^{\lessgtr}\left(\mathbf{k}^{\prime}\mathbf{,}\omega_{1}+\omega_{2}-\omega\right).
\end{eqnarray}
$N$ combines the valley and spin degrees of freedom and takes the
value $N=4$. Next we use the fact that within a quasiparticle description
the upper and lower propagators are expressed in terms of the distribution
functions $f_{\lambda}\left(\mathbf{k},\mathbf{r},t\right)$ as
\begin{eqnarray}
G_{\lambda}^{>}\left(\mathbf{k,}\mathbf{r},\omega,t\right) & = & -i2\pi\delta\left(\omega-\varepsilon_{\lambda}\left(\mathbf{k}\right)\right)\left(1-f_{\lambda}\left(\mathbf{k},\mathbf{r},t\right)\right)\nonumber \\
G_{\lambda}^{<}\left(\mathbf{k,}\mathbf{r},\omega,t\right) & = & i2\pi\delta\left(\omega-\varepsilon_{\lambda}\left(\mathbf{k}\right)\right)f_{\lambda}\left(\mathbf{k},\mathbf{r},t\right)
\end{eqnarray}
As usual, $\mathbf{k}$ and $\omega$ stand for the Fourier-transformed
variables of the relative coordinates and times while $\mathbf{r}$
and $t$ stand for the center of gravity or mean time. 

The collision operator can now we determined from the self energies
$\Sigma^{<}$ and $\Sigma^{>}$: 
\begin{equation}
{\cal C}_{\lambda}\left(\mathbf{k}\right)=-i\Sigma_{\lambda}^{<}\left(\mathbf{k,}\varepsilon_{\lambda}\left(\mathbf{k}\right)\right)\left(1-f_{\lambda}\left(\mathbf{k}\right)\right)-i\Sigma_{\lambda}^{>}\left(\mathbf{k,}\varepsilon_{\lambda}\left(\mathbf{k}\right)\right)f_{\lambda}\left(\mathbf{k}\right).
\end{equation}
For simplicity we only keep the momentum $\mathbf{k}$ and band index
$\lambda$. Inserting $G^{>}$ and $G^{<}$ into the self energies
yields with the linearization Eq.(6) of the main paper the result
for the collision operator given in Eq.(8) of the main paper. The
matrix elements $\gamma_{\mathbf{k},\mathbf{k}',\mathbf{q}}^{\left(1,2\right)}$
of that equation are given as:

\begin{eqnarray}
\gamma_{1}\left(\mathbf{k,k}^{\prime},\mathbf{q}\right) & = & \left(N-1\right)\left\vert T_{A}\left(\mathbf{k,k}^{\prime},\mathbf{q}\right)\right\vert ^{2}+\frac{1}{2}\left\vert T_{A}\left(\mathbf{k,k}^{\prime},\mathbf{k}^{\prime}-\mathbf{q-k}\right)-T_{A}\left(\mathbf{k,k}^{\prime},\mathbf{q}\right)\right\vert ^{2}\nonumber \\
 &  & -\left\vert T_{A}\left(\mathbf{k,k}^{\prime},\mathbf{k}^{\prime}-\mathbf{q-k}\right)\right\vert ^{2}\nonumber \\
\gamma_{2}\left(\mathbf{k,k}^{\prime},\mathbf{q}\right) & = & \left(N-1\right)\left\vert T_{B}\left(\mathbf{k,\mathbf{k}^{\prime}},\mathbf{\mathbf{k}^{\prime}-k-\mathbf{q}}\right)\right\vert ^{2}+\left(N-1\right)\left\vert T_{A}\left(\mathbf{k,\mathbf{k}^{\prime}},\mathbf{q}\right)\right\vert ^{2}\nonumber \\
 &  & +\left\vert T_{A}\left(\mathbf{k,\mathbf{k}^{\prime}},\mathbf{q}\right)-T_{B}\left(\mathbf{k,\mathbf{k}^{\prime}},\mathbf{\mathbf{k}^{\prime}-\mathbf{q}}\mathbf{-k}\right)\right\vert ^{2},
\end{eqnarray}
with 
\begin{eqnarray*}
T_{A}\left(\mathbf{k,k}^{\prime},\mathbf{q}\right) & = & T_{++++}\left(\mathbf{k,k}^{\prime},\mathbf{q}\right)=T_{----}\left(\mathbf{k,k}^{\prime},\mathbf{q}\right)\\
 & = & T_{+--+}\left(\mathbf{k,k}^{\prime},\mathbf{q}\right)=T_{-++-}\left(\mathbf{k,k}^{\prime},\mathbf{q}\right)\\
 & = & \frac{V\left(q\right)}{4}\left(1+\frac{\left(K+Q\right)K^{\ast}}{\left\vert \mathbf{k+q}\right\vert k}\right)\left(1+\frac{\left(K^{\prime}-Q\right)K^{\prime\ast}}{\left\vert \mathbf{k}^{\prime}\mathbf{-q}\right\vert k^{\prime}}\right)
\end{eqnarray*}
and 
\begin{eqnarray}
T_{B}\left(\mathbf{k,k}^{\prime},\mathbf{q}\right) & = & T_{++--}\left(\mathbf{k,k}^{\prime},\mathbf{q}\right)=T_{--++}\left(\mathbf{k,k}^{\prime},\mathbf{q}\right)\nonumber \\
 & = & \frac{V\left(q\right)}{4}\left(1-\frac{\left(K+Q\right)K^{\ast}}{\left\vert \mathbf{k+q}\right\vert k}\right)\left(1-\frac{\left(K^{\prime}-Q\right)K^{\prime\ast}}{\left\vert \mathbf{k}^{\prime}\mathbf{-q}\right\vert k^{\prime}}\right)
\end{eqnarray}
Upper-case letters like $K=k_{x}+ik_{y}$ etc. combine the two components
of the momentum onto a complex variable.

From the same unitary transformation also follows that 
\begin{equation}
U_{\mathbf{k}}\mathbf{\sigma}U_{\mathbf{k}}^{-1}=\frac{\mathbf{k}}{k}\sigma_{z}-\frac{\mathbf{k\times e}_{z}}{k}\sigma_{y}.
\end{equation}
This can be used to analyze the current 
\begin{equation}
\mathbf{j=}ev\int_{\mathbf{k}}\psi^{\dagger}\left(\mathbf{k}\right)\mathbf{\sigma}\psi\left(\mathbf{k}\right)
\end{equation}
of Dirac particles which consists of intra- and inter-band contributions:
\begin{equation}
\mathbf{j=j}_{\mathrm{intra}}+\mathbf{j}_{\mathrm{inter}}.
\end{equation}
The two terms are given as 
\begin{eqnarray}
\mathbf{j}_{\mathrm{intra}} & = & ev\int_{\mathbf{k}}\sum_{\lambda=\pm}\frac{\lambda\mathbf{k}}{k}\gamma_{\mathbf{k},\lambda}^{\dagger}\gamma_{\mathbf{k},\lambda}\nonumber \\
\mathbf{j}_{\mathrm{inter}} & = & iev\int_{\mathbf{k}}\frac{\mathbf{k\times e}_{z}}{k}\left(\gamma_{\mathbf{k},+}^{\dagger}\gamma_{\mathbf{k},-}-\gamma_{\mathbf{k},-}^{\dagger}\gamma_{\mathbf{k},+}\right).
\end{eqnarray}
Thus, the velocity used in our Eq. (4) of the main paper is precisely
the expression $\mathbf{v}_{\mathbf{k}\lambda}=v\frac{\lambda\mathbf{k}}{k}$
of the intraband current $\mathbf{j}_{\mathrm{intra}}$. Spin-momentum
locking is included naturally, if one goes to the helical states of
the upper and lower Dirac cone. The hydrodynamic response is governed
by strong collisions of intraband excitations.

\section{Identification of the collinear modes at finite angular momentum}

In this section we determine the zero modes of the collision operator
of Eq.(8) if we confine ourselves to collinear collision processes.
To this end we need to find under what conditions the two expressions
\begin{eqnarray}
A_{\mathbf{k},\mathbf{k}',\mathbf{q},\lambda}^{\left(1\right)} & = & \psi_{\mathbf{k}+\mathbf{q}\lambda}+\psi_{\mathbf{k}'-\mathbf{q}\lambda}-\psi_{\mathbf{k}'\lambda}-\psi_{\mathbf{k}\lambda}\nonumber \\
A_{\mathbf{k},\mathbf{k}',\mathbf{q},\lambda}^{\left(2\right)} & = & \psi_{\mathbf{k}+\mathbf{q}\lambda}-\psi_{-\mathbf{k}'+\mathbf{q}\bar{\lambda}}+\psi_{-\mathbf{k}'\bar{\lambda}}-\psi_{\mathbf{k}\lambda},
\end{eqnarray}
that occur in Eq.(8), vanish. Here we have to include the additional
constrain
\begin{equation}
\left|\mathbf{k}+\mathbf{q}\right|+\left|\mathbf{k}'-\mathbf{q}\right|=\left|\mathbf{k}\right|+\left|\mathbf{k}'\right|\label{eq:energy}
\end{equation}
that follows from energy conservation. 

By collinear modes we mean that all involved momenta are either parallel
or antiparallel. As discussed in Ref.\cite{Fritz2008-1} we consider
such zero modes of collinear processes because all other processes
are suppressed by $1/\log\left(1/\alpha\right)$ where $\alpha$ is
the fine-structure constant. Of course, the analysis allows for scattering
processes that are not collinear; the issue is merely that distribution
functions that become zero modes for collinear scattering are enhanced
relative to those that are no such zero modes. Finally we comment
that the main conclusion of our paper, namely that the scattering
rate depends on angular momentum like $\tau_{m}^{-1}\propto k_{B}T\left|m\right|$,
is unchanged if we go beyond the collinear mode regime.

One immediately finds that $A^{\left(1\right)}=B^{\left(1\right)}=0$
subject to Eq.\ref{eq:energy} is obeyed by $\psi_{\mathbf{k},\lambda}=1$,
$\psi_{\mathbf{k},\lambda}=\mathbf{k}$, and $\psi_{\mathbf{k},\lambda}=\lambda\left|\mathbf{k}\right|$,
regardless of whether we confine ourselves to collinear modes. These
modes correspond to the conservation of charge, momentum, and energy,
respectively. In addition to these modes one also finds $\psi_{\mathbf{k},\lambda}=\lambda$
is a zero mode. It corresponds to the fact that second order perturbation
theory does not relax a charge imbalance between the upper and lower
Dirac cone.

Next we consider distribution functions 
\begin{equation}
\psi_{\mathbf{k},\lambda}=a_{\lambda,m}\left(k\right)e^{im\theta_{\mathbf{k}}},
\end{equation}
where $k=\left|\mathbf{k}\right|$ is the magnitude of the momentum
and $\theta_{\mathbf{k}}$ its polar angle: $\mathbf{k}=k\left(\cos\theta_{\mathbf{k}},\sin\theta_{\mathbf{k}}\right)$.
Collinear scattering corresponds to 
\begin{equation}
\mathbf{\theta_{\mathbf{k}}=\theta_{\mathbf{k}'}+s_{1}\pi=\theta_{\mathbf{k}+\mathbf{q}}+s_{2}\pi=\theta_{\mathbf{k}'-\mathbf{q}}+s_{3}\pi},
\end{equation}
where even or odd $s_{i}$ correspond to parallel and antiparallel
momenta relative to $\mathbf{k}$. We first show that all $s_{i}$
are even due to energy conservation. To this end we assume without
restriction that $\mathbf{k}=\left(k,0\right)$ with $k>0$. Then
$\mathbf{k}'=u\left(k,0\right)$ and $\mathbf{q}=w\left(k,0\right)$,
where we do not assume that $u$ and $w$ are positive. Energy conservation
now implies 
\begin{equation}
1+\left|u\right|=\left|1+w\right|+\left|u-w\right|.
\end{equation}
We need to fulfill this condition for an extended set of variables,
not just for an isolated point in momentum space. This implies that
$1+w>0$ so we can cancel the ``$1$'' on both sides. Then, to be
able to cancel $w$ it must hold that $u>w$, which in turn implies
$u>0$ to cancel $u$. Thus, we find that the momenta $\mathbf{k}'$,
$\mathbf{k}+\mathbf{q}$, and $\mathbf{k}'-\mathbf{q}$ point in the
same direction as $\mathbf{k}$ even though $\mathbf{q}$ is allowed
to point in the opposite direction. It follows that we can assume
without restriction that 
\begin{equation}
\mathbf{\theta_{\mathbf{k}}=\theta_{\mathbf{k}'}=\theta_{\mathbf{k}+\mathbf{q}}=\theta_{\mathbf{k}'-\mathbf{q}}}.
\end{equation}
If we use that $e^{im\theta_{-\mathbf{p}}}=\left(-1\right)^{m}e^{im\theta_{\mathbf{p}}}$we
obtain
\begin{eqnarray}
A_{\mathbf{k},\mathbf{k}',\mathbf{q},\lambda}^{\left(1\right)} & = & \left(a_{\lambda,m}\left(\left|\mathbf{k}+\mathbf{q}\right|\right)+a_{\lambda,m}\left(\left|\mathbf{k}'-\mathbf{q}\right|\right)-a_{\lambda,m}\left(k'\right)-a_{\lambda,m}\left(k\right)\right)e^{im\theta_{\mathbf{k}}}\nonumber \\
A_{\mathbf{k},\mathbf{k}',\mathbf{q},\lambda}^{\left(2\right)} & = & \left(a_{\lambda,m}\left(\left|\mathbf{k}+\mathbf{q}\right|\right)-\left(-1\right)^{m}a_{\bar{\lambda},m}\left(\left|\mathbf{k}'-\mathbf{q}\right|\right)+\left(-1\right)^{m}a_{\bar{\lambda},m}\left(k'\right)-a_{\lambda,m}\left(k\right)\right)e^{im\theta_{\mathbf{k}}}.
\end{eqnarray}
It is now easy so find that there are the following solutions that
yield $A^{\left(1\right)}=B^{\left(1\right)}=0$ subject to Eq.\ref{eq:energy}:
\begin{eqnarray}
a_{\lambda,m}\left(k\right) & = & 1\nonumber \\
a_{\lambda,m}\left(k\right) & = & \lambda.
\end{eqnarray}
In addition one finds $a_{\lambda,m}\left(k\right)=\lambda\left|\mathbf{k}\right|$
if $m$ is even and $a_{\lambda,m}\left(k\right)=\left|\mathbf{k}\right|$
if $m$ is odd. 

Thus, we can write that the following modes are zero modes in the
collinear scattering limit 
\begin{equation}
\psi_{\mathbf{k},\lambda}=\lambda^{m}e^{im\theta_{\mathbf{k}}}\left(1,\lambda,\lambda\left|\mathbf{k}\right|\right)\label{eq:collinear_zero_modes}
\end{equation}
which is Eq.(9) of the main paper.

\section{Superdiffusion and non-local transport}

\begin{figure}
\centering{}\includegraphics[scale=0.4]{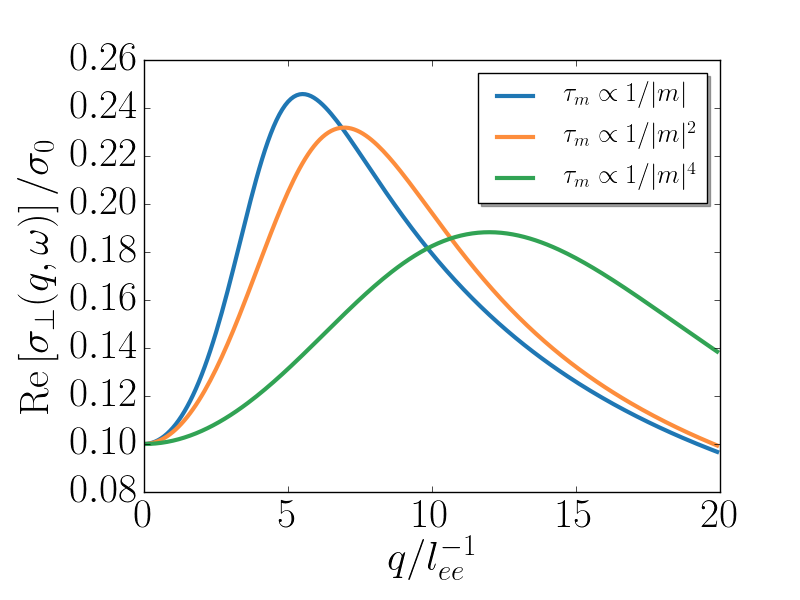}\caption{Real part of the transverse conductivity $\sigma_{\bot}\left(\omega,q\right)$
for different dependecies of the scattering times $\tau_{m}$ on $m$.
This result was obtain by solving Eq. (\ref{eq:Boltzmann_collinear})
numerically.\label{fig:Cond_trans}}
\end{figure}

The Fokker-Planck equation (2) of the main text
\begin{equation}
\left(\partial_{t}+\mathbf{v}_{\mathbf{k}\lambda}\cdot\nabla_{\mathbf{x}}-\tau_{L}^{-1}\left(\frac{\partial^{2}}{\partial\theta^{2}}\right)^{1/2}\right)f_{\mathbf{k}\lambda}=S_{\mathbf{k}\lambda}\label{eq:Boltzmann_supplement}
\end{equation}
 can be used to calculate the response of graphene electrons to an
external perturbation, such as for example an electric field. In this
case the force term is given by 
\[
S_{\mathbf{k}\lambda}=-e\mathbf{E}\left(\mathbf{q},\omega\right)\cdot\frac{\partial f_{\mathbf{k}\lambda}}{\partial\mathbf{k}},
\]
where $\mathbf{E\left(\mathbf{q},\omega\right)}=E_{0}e^{i\left(\mathbf{q}\cdot\mathbf{x}-\omega t\right)}\mathbf{e}_{x}$
is the electric field. To first order in the electric field it is
\[
S_{\mathbf{k}\lambda}=-eE_{0}e^{i\mathbf{q}\cdot\mathbf{x}}\cos\theta\left(\lambda\hbar v\beta\right)f_{k}^{\left(0\right)}\left(1-f_{k}^{\left(0\right)}\right).
\]
We perform a Fourier transform $t\rightarrow\omega$, $\mathbf{x}\rightarrow\mathbf{q}$
and project the equation (\ref{eq:Boltzmann_supplement}) onto the
collinear zero modes (\ref{eq:collinear_zero_modes}) using the scalar
product $\Braket{\phi|\chi}=\sum_{\lambda}\int\frac{d^{2}k}{\left(2\pi\right)^{2}}\phi_{\bm{k},\lambda}\chi_{\bm{k},\lambda}$.
The result is a simplified version of the Boltzmann equation:
\begin{equation}
-i\omega\delta_{m,m'}+\frac{1}{2}ivq\left(e^{-i\vartheta_{\bm{q}}}\delta_{m,m'+1}+e^{i\vartheta_{\bm{q}}}\delta_{m,m'-1}\right)+\frac{1}{\tau_{L}\left|m\right|}=-\frac{1}{2}eE_{0}v\beta\delta_{\left|m\right|,1},\label{eq:Boltzmann_collinear}
\end{equation}
where $m$ labels the angular harmonic of the collinear zero mode
and $\vartheta_{\bm{q}}$ is the angle of the wave vector $\mathbf{q}$
with respect to the $x$-axis. This equation is exact in the limit
of a small coupling constant $\alpha$, where collinear events dominate
the electron-electron scattering \cite{Fritz2008-1}. Notice, that
the electric field only couples to angular harmonics with $\left|m\right|=1$,
however for a spatially inhomogeneous field with $q\neq0$, the second
right hand side term of Eq. (\ref{eq:Boltzmann_collinear}) couples
all angular harmonics. Therefore, information on the $m$-dependence
of the scattering times can be extracted from the non-local (i.e.
$\mathbf{q}$-dependent) electric conductivity $\sigma_{\alpha\beta}\left(\mathbf{q},\omega\right)$,
which is defined through
\[
j_{\alpha}\left(\mathbf{q},\omega\right)=\sigma_{\alpha\beta}\left(\mathbf{q},\omega\right)E\left(\mathbf{q},\omega\right).
\]
The conductivity tensor $\sigma_{\alpha\beta}\left(\mathbf{q},\omega\right)$
can be decomposed into longitudinal and transverse parts according
to
\[
\sigma_{\alpha\beta}=\frac{q_{\alpha}q_{\beta}}{q^{2}}\sigma_{\parallel}\left(\omega,q\right)+\left(\delta_{\alpha\beta}-\frac{q_{\alpha}q_{\beta}}{q^{2}}\right)\sigma_{\bot}\left(\omega,q\right),
\]
where $\sigma_{\parallel/\bot}\left(\omega,q\right)$ only depend
on the magnitude of $\mathbf{q}$. Fig \ref{fig:Cond_trans} shows
the influence of the $m$-dependence of the scattering time $\tau_{m}$
on the real part of the transverse conductivity $\sigma_{\bot}\left(\omega,q\right)$.
We conclude that the non-local conductivity, playing an important
role in experiments on surface acoustic waves \cite{Wixforth1986,Simon1996},
provides a possibility to detect the peculiar dependence of the scattering
times $\tau_{m}$ on $m$, and to confirm the Lévy flight behavior
predicted in the main text.
For completeness, we mention that the Boltzmann equation (\ref{eq:Boltzmann_collinear})
can be solved exactly, and the expressions for the non-local conductivities
can be written down in closed form \cite{EK2019}:
\begin{eqnarray}
\sigma_{\parallel}\left(q,\omega\right) & = & \frac{\sigma_{0}}{1+i\tau_{1}\omega-\frac{1}{4}v^{2}\tau_{1}q^{2}\left(\frac{2i}{\omega}-\frac{1}{M\left(q,\omega\right)+i\omega}\right)},\nonumber \\
\sigma_{\bot}\left(q,\omega\right) & = & \frac{\sigma_{0}}{1+i\tau_{1}\omega+\frac{\frac{1}{4}v^{2}\tau_{1}q^{2}}{M\left(q,\omega\right)+i\omega}}.\label{eq:charge_conductivity_result}
\end{eqnarray}
Here, $\sigma_{0}=N\frac{e^{2}\log\left(2\right)\tau_{\sigma,1}}{2\pi\beta\hbar^{2}}$
is the quantum critical conductivity calculated in Ref. \cite{Fritz2008-1}
and $M\left(q,\omega\right)$ is the memory function summerizing the
effects of higher angular harmonics: 
\[
M\left(q,\omega\right)=\tau_{2}^{-1}+\frac{1}{2}vq\frac{\textrm{I}_{3+i\omega\tau_{L}}\left(\tau_{L}vq\right)}{\textrm{I}_{2+i\omega\tau_{L}}\left(\tau_{L}vq\right)},
\]
where $I_{\nu}\left(z\right)$ is the modified Bessel function.


\begin{thebibliography}{10}
\bibitem{Chandrasekhar1943}S. Chandrasekhar, \emph{Stochastic Problems
in Physics and Astronomy}, Rev. Mod. Phys. 15, 1 (1943).

\bibitem{Rosenbluth1957}M. N. Rosenbluth, W. M. MacDonald, and D.
L. Judd, \emph{Fokker-Planck Equation for an Inverse-Square Force},
Phys. Rev. \textbf{107} 1 (1957).

\bibitem{Chernoff1990}D. F. Chernoff and M. D. Weinberg, Evolution
of globular clusters in the Galaxy, Astrophysical Journal \textbf{351},121
(1990).

\bibitem{Sachdevbook}S. Sachdev, \emph{Quantum Phase Transitions}
(Cambridge Univ. Press 1999).

\bibitem{Hertz1976}J. A. Hertz, \emph{Quantum critical phenomena},
Phys. Rev. B \textbf{14}, 1165 (1976).

\bibitem{Moriya1984}T. Moriya, \emph{Spin Fluctuations in Itinerant
Electron Magnetism}, Springer Series in Solid-State Science 52, Springer,
Berlin-Heidelberg (1984).

\bibitem{Millis1993}A. J. Millis, \emph{Effect of a nonzero temperature
on quantum critical points in itinerant fermion systems,} Phys. Rev.
B \textbf{48}, 7183 (1993).

\bibitem{Laughlin2001}R. B. Laughlin, G. G. Lonzarich, P. Monthoux,
and D. Pines, \emph{The Quantum Criticality Conundrum}, Adv. Phys.
50, 361 (2001).

\bibitem{Millis2002}A. J. Millis, A. J. Schofield, G. G. Lonzarich,
and S. A. Grigera, \emph{Metamagnetic Quantum Criticality}, Phys.
Rev. Lett. 88, 217204 (2002).

\bibitem{Abanov2003}Ar. Abanov, A. V. Chubukov, and J. Schmalian,
\emph{Quantum-critical theory of the spin-fermion model and its application
to cuprates: Normal state analysis}, Adv. Phys. \textbf{52}, 119 (2003).

\bibitem{Loehneysen2007}H. v. Löhneysen, A. Rosch, M. Vojta, and
P. Wölfle, \emph{Fermi-liquid instabilities at magnetic quantum phase
transitions}, Rev. Mod. Phys. \textbf{79}, 1015 (2007).

\bibitem{DellAnna2007}L. Dell\textquoteright Anna and W. Metzner,
Electrical Resistivity near Pomeranchuk Instability in Two Dimensions,
Phys. Rev. Lett. 98, 136402 (2007).

\bibitem{Metlitski2010}M. A. Metlitski and S. Sachdev, Quantum phase
transitions of metals in two spatial dimensions. I. Ising-nematic
order, Phys. Rev. B \textbf{82}, 075127 (2010).

\bibitem{Schattner2016}Y. Schattner, S. Lederer, S. A. Kivelson,
and E. Berg, \emph{Ising Nematic Quantum Critical Point in a Metal:
A Monte Carlo Study}, Phys. Rev. X \textbf{6}, 031028 (2016).

\bibitem{Damle1997}K. Damle and S. Sachdev, \emph{Nonzero-temperature
transport near quantum critical points,} Phys. Rev. B \textbf{56},
8714 (1997).

\bibitem{Sheehy2007}Daniel E. Sheehy and Jörg Schmalian, Quantum
Critical Scaling in Graphene, Phys. Rev. Lett. 99, 226803 (2007).

\bibitem{Gurzhi1995}R. N. Gurzhi, A. N. Kalinenko, and A. I. Kopeliovich
Phys. Rev. Lett. \textbf{74}, 3872 (1995).

\bibitem{Gurzhi1996}R.N. Gurzhi, A.N. Kalinenko, A.I. Kopeliovich,
Surface Science \textbf{361}/\textbf{362, }497 (1996).

\bibitem{Maslov2011}D. L. Maslov, V. I. Yudson, and A. V. Chubukov,
Phys. Rev. Lett. \textbf{106}, 106403 (2011).

\bibitem{Pal2012}H. K. Pal, V. I. Yudson, and D. L. Maslov, \emph{Resistivity
of non-Galileian invariant Fermi- and Non-Fermi liquids}, Lithuanian
Journal of Physics, \textbf{52}, 142 (2012).

\bibitem{Maslov2017}D. L. Maslov and A. V. Chubukov, Optical response
of correlated electron systems, Rep. Prog. Phys. \textbf{80}, 026503
(2017).

\bibitem{Ledwith2019}P. J. Ledwith, H. Guo, L. Levitov, \emph{The
Hierarchy of Excitation Lifetimes in Two-Dimensional Fermi Gases},
arXiv:1905.03751.

\bibitem{Fritz2008}L. Fritz, J. Schmalian, M. Müller and S. Sachdev,
\emph{Quantum critical transport in clean graphene}, Phys. Rev. B
\textbf{78}, 085416 (2008).

\bibitem{Kashuba2008}A. B. Kashuba, \emph{Conductivity of defectless
graphene}, Phys. Rev. B 78, 085415 (2008).

\bibitem{Mueller2009}M. Müller, J. Schmalian and L. Fritz, \emph{Graphene:
A Nearly Perfect Fluid}, Phys. Rev. Lett. \textbf{103}, 025301 (2009).

\bibitem{Schuett2011}M. Schütt, P. M. Ostrovsky, I. V. Gornyi, and
A. D. Mirlin, \emph{Coulomb interaction in graphene: Relaxation rates
and transport}, Phys. Rev. B 83, 155441 (2011).

\bibitem{Kiselev2019}E. I. Kiselev, J. Schmalian, \emph{Boundary
conditions of viscous electron flow}, Phys. Rev. B \textbf{99}, 035430
(2019).

\bibitem{Levy1954}P. Lévy, P. \emph{Théorie de l'Addition des Variables
Aléatoires} Gauthier-Villars, Paris, (1954).

\bibitem{Mandelbrot1977}B. Mandelbrot, T\emph{he Fractal Geometry
of Nature}, Freeman, New York, (1977).

\bibitem{Feller1972}W. Feller, \emph{An Introduction to Probability
Theory and its Applications}, 2nd ed., Vol. 2, Wiley, New York, (1971).

\bibitem{Samko1993}S. G. Samko, A. A. Kilbas and O. I. Marichev,
\emph{Fractional Integrals and Derivatives, Theory and Applications,}
Gordon and Breach, Amsterdam, (1993).

\bibitem{Hermann2014}R. Hermann, \emph{Fractional Calculus: An Introduction
for Physicists} (2nd ed.). New Jersey: World Scientific Publishing
(2014).

\bibitem{Bartumeus2005}F. Bartumeus, M. G. E. Da Luz, G. M. Viswanathan,
J. Catalan, \emph{Animal search strategies: A quantitative random-walk
analysis}. Ecology \textbf{86}, 3078 (2005).

\bibitem{Reynolds2009}A. M. Reynolds and C. J. Rhodes, \emph{The
Lévy flight paradigm: random search patterns and mechanisms, }Ecology,
90,, 877 (2009).

\bibitem{Mantegna1997}R. N. Mantegna and H. E. Stanley, \emph{Econophysics:
Scaling and its breakdown in finance. Journal of statistical Physics},
\textbf{89}, 469 (1997).

\bibitem{Corral2006}A. Corral, \emph{Universal earthquake-occurrence
jumps, correlations with time, and anomalous diffusion} Phys. Rev.
Lett. \textbf{97}, 178501 (2006).

\bibitem{Levy1939}P. Levy, \emph{L\textquoteright addition des variables
aléatoires définies sur une circonférence,} Bulletin de la Societe
mathematique de France, 67, 1 (1939),

\bibitem{Mardia1999}Kanti V. Mardia and Peter E. Jupp, \emph{Directional
Statistics}, Wiley (1999) ISBN 978-0-471-95333-3. 

\bibitem{supplementary material}see supplementary material.

\bibitem{Gallagher2019}P. Gallagher, C.-S. Yang, T. Lyu, F. Tian,
R. Kou, H. Zhang, K. Watanabe, T. Taniguchi, and F. Wang, \emph{Quantum-critical
conductivity of the Dirac fluid in graphene, }Science \textbf{364},
158 (2019).

\bibitem{Gattenloehner2016}S. Gattenlöhner, I. V. Gornyi, P. M. Ostrovsky,
B. Trauzettel, A. D. Mirlin, M. Titov, \textit{Lévy flights due to
anisotropic disorder in graphene}. Phys. Rev. Lett. \textbf{117},
046603 (2016).

\bibitem{Briskot2014}U. Briskot, I. A. Dmitriev, and A. D. Mirlin,
\emph{Relaxation of optically excited carriers in graphene: Anomalous
diffusion and Lévy flights}, Phys. Rev. B \textbf{89}, 075414 (2014).

\bibitem{Barthelemy2008}P. Barthelemy, J. Bertolotti, and D. S. Wiersma,
\emph{A Lévy flight for light, }Nature \textbf{453}, 495 (2008).

\bibitem{Crossno2016}J. Crossno, J. K. Shi, K. Wang, X. Liu, A. Harzheim,
A. Lucas, S. Sachdev, P. Kim, T. Taniguchi, K. Watanabe, T. A. Ohki,
K. C. Fong, \emph{Observation of the Dirac fluid and the breakdown
of the Wiedemann-Franz law in graphene}, Science \textbf{351}, 1058
(2016).

\bibitem{Predel2000}H. Predel, H. Buhmann, L. W. Molenkamp, R. N.
Gurzhi, A. N. Kalinenko, A. I. Kopeliovich, A. V. Yanovsky, Phys.
Rev. B \textbf{62}, 2057 (2000).

\bibitem{Wang2019}K. Wang, M. M. Elahi, L. Wang, K. M. M. Habib,
T. Taniguchi, K. Watanabe, J. Hone, A. W. Ghosh, G.-H. Lee, P. Kim,
PNAS \textbf{116}, 6575 (2019).

\bibitem{Mirlin1997}A. D. Mirlin and P. Wölfle, \textit{Composite
fermions in the fractional quantum Hall effect: Transport at finite
wave vector}, Phys. Rev. Lett. \textbf{78}, 371 (1997).

\bibitem{Kiselev2019b}E. I. Kiselev and J. Schmalian, unpublished.Supplementary
material: Lévy flights and hydrodynamic superdiffusion on the Dirac
cone of graphene

\end{thebibliography}

\begin{thebibliography}{10}
\bibitem{Chechkin2004}A. V. Chechkin, V. Y. Gonchar, J. Klafter,
R. Metzler, L. V. Tanatarov, J. Stat. Phys. \textbf{115}, 1505 (2004).

\bibitem{Metzler2012}R. Metzler, A. V. Chechkin, J. Klafter, Levy
Statistics and Anomalous Transport, Computational Complexity, Springer,
2012.

\bibitem{Kurz2014}G. Kurz, I. Gilitschenski and U. D. Hanebeck, \textquotedbl{}Efficient
evaluation of the probability density function of a wrapped normal
distribution\textquotedbl{} in 2014 Sensor Data Fusion: Trends, Solutions,
Applications (SDF), IEEE (2014).

\bibitem{Fritz2008-1}L. Fritz, J. Schmalian, M. Müller and S. Sachdev,
\emph{Quantum critical transport in clean graphene}, Phys. Rev. B
\textbf{78}, 085416 (2008).

\bibitem{Wixforth1986}A. Wixforth, J. P. Kotthaus, and G. Weimann
Phys. Rev. Lett. \textbf{56}, 2104 (1986).

\bibitem{Simon1996}H. S. Simon, Phys. Rev. B \textbf{54}, 13878 (1996).

\bibitem{EK2019}E. I. Kiselev, J. Schmalian, unpublished.\end{thebibliography}
\end{document}